# Reconfigurable Microwave Photonic All-Pass Filters based on Optical Kerr Microcombs

David J. Moss*

*Optical Sciences Center, Swinburne University of Technology, Hawthorn, VIC 3122, Australia;*
E-mail: dmoss@swin.edu.au

■ **ABSTRACT:**

Microwave photonic (MWP) all-pass filters (APFs), which enable phase manipulation without altering signal amplitudes, are essential for phase engineering in high-bandwidth microwave signal processing. However, achieving a high degree of reconfigurability in the filter order and characteristic parameters of APFs remains challenging for conventional methods. Here, we propose and experimentally demonstrate a new strategy for realizing MWP APFs based on a microcomb-driven transversal filter system, enabling high reconfigurability, sharp phase transitions with high group delays, and low loss. We experimentally demonstrate both 1$^{st}$- and 2$^{nd}$-order APFs, achieving a maximum amplitude variation (MAV) of ~1.95 dB and a maximum group delay (MGD) of ~580 ps for the former, and a MAV of ~2.01 dB and an MGD of ~1740 ps for the latter. In addition, both independent and joint tuning of APF characteristic parameters are demonstrated by simply programming the tap coefficients without changing any hardware, enabling highly reconfigurable filter response for diverse phase-engineering functionalities. These results highlight the potential of our approach for implementing MWP APFs with both high performance and reconfigurability for practical applications requiring dynamic and versatile phase control.

## ■ INTRODUCTION

Microwave photonic (MWP) filters, which harness photonic technologies for microwave signal filtering, have found wide applications for modern communications, radar, and high-speed signal processing [1-4]. Compared to conventional electrical microwave filters, MWP filters offer several attractive advantages, including broad operation bandwidths beyond the intrinsic limits of electronic devices, low loss for improved signal-to-noise ratio, and strong immunity to electromagnetic interference [5-7]. As a fundamental class of MWP filters, MWP all-pass filters (APFs) feature a flat amplitude response and a frequency-dependent phase response, enabling phase manipulation without altering signal amplitudes. This capability has made them attractive for phase-engineering applications such as phased arrays [8], phase shifters [9, 10], true time delay lines [11, 12], and beamforming [13, 14].

In principle, MWP APFs can be realized by direct mapping the optical response of a lossless micro-ring resonator (MRR) to the microwave domain [15], where the device transfer function satisfies the all-pass condition, with its zeros located at the inverse-conjugate positions of the poles [16]. However, practical waveguide propagation loss [17, 18] prevents the strict satisfaction of the ideal all-pass condition, resulting in amplitude distortion in the filter response. To address this, several schemes have been proposed, including direct loss compensation using optical amplifiers [19], self-compensation strategies [20], and passive compensation based on feedback waveguides [21, 22]. However, these approaches typically rely on accurate gain compensation, critical-coupling matching, or additional interference engineering, which could introduce additional design complexity and insertion loss. In addition, the characteristic parameters of APFs, such as the center frequency and phase damping factor, are usually governed by multiple structural parameters of the passive optical filters, restricting their reconfigurability to minor adjustments via PN junctions [23, 24] or thermo-optic heaters [25, 26].

In this work, we propose and experimentally demonstrate MWP APFs based on a microcomb-driven transversal filter system, featuring a high degree of reconfigurability, sharp phase transitions with high group delays, and low loss. In our experiments, we demonstrate both 1st- and 2nd-order APFs, with the former achieving a maximum amplitude variation (MAV) of ~1.95 dB and a maximum group delay (MGD) of ~580 ps, and the latter achieving a MAV of ~2.01 dB and an MGD of ~1740 ps. Moreover, by simply programming the tap coefficients without changing any hardware, we demonstrate both independent and joint tuning of APF characteristic parameters to realize three typical phase-engineering functionalities. These results validate the effectiveness of our approach for implementing MWP APFs that combine excellent performance with high reconfigurability for practical applications requiring dynamic and versatile phase control.

## ■ OPERATION PRINCIPLE

APFs feature frequency-dependent phase response and frequency-independent amplitude response, typically realized by arranging the poles and zeros in reciprocal conjugate pairs [27]. In this work, we focus on 1st- and 2nd-order APFs, which constitute the fundamental types of APFs. The transfer function of a 1st-order APF can be expressed as [28]

$$H(s) = \frac{s - \omega_c}{s + \omega_c}, \tag{1}$$

where $s$ is the complex frequency and $\omega_c = 2\pi f_c$ is the center angular frequency, with $f_c$ denoting the center frequency. For practical implementation of microwave APFs, the transfer function in **Eq. (1)** is evaluated along the imaginary axis by setting $s = j2\pi f$, with $f$ denoting the microwave frequency. Therefore, **Eq. (1)** can be rewritten as

$$H(f) = \frac{jf - f_c}{jf + f_c}. \tag{2}$$

A 2nd-order APF can be realized by cascading two 1st-order APFs, and its general transfer function can be given by [29]

$$H(s) = \frac{s^2 - 2\xi\omega_c s + \omega_c^2}{s^2 + 2\xi\omega_c s + \omega_c^2}, \tag{3}$$

where $\xi$ is the phase damping factor that determines the normalized pole-zero locations in the complex plane, thereby controlling the roll-off of the phase transition around the center frequency $f_c$. Similarly, substituting $s = j2\pi f$ and $\omega_c = 2\pi f_c$ into **Eq. (3)** yields

$$H(f) = \frac{f_c^2 - f^2 - j2\xi f_c f}{f_c^2 - f^2 + j2\xi f_c f}, \tag{4}$$

which is the spectral transfer function of 2nd-order microwave APFs. Compared with the spectral transfer function of 1st-order APFs in **Eq. (2)**, which has a single characteristic parameter $f_c$, the spectral transfer function of 2nd-order APFs in **Eq. (4)** has two characteristic parameters, $f_c$ and $\xi$, thereby allowing a higher degree of freedom in phase response engineering.

To quantify the roll-off of the phase transition for APFs, the group delay response, calculated by taking the negative derivative of the phase response in **Eq. (3)** with respect to the angular frequency $\omega$, can be given by [30]

$$T_{gd}(\omega) = \frac{4\xi\omega_c(\omega_c^2 + \omega^2)}{(\omega_c^2 - \omega^2)^2 + 4\xi^2\omega^2\omega_c^2}, \tag{5}$$

By substituting $\omega = 2\pi f$ and $\omega_c = 2\pi f_c$ into **Eq. (5),** the group delay response of microwave APFs can be expressed as

$$T_{gd}(f) = \frac{2\xi f_c(f_c^2 + f^2)}{\pi\left[(f_c^2 - f^2)^2 + 4\xi^2 f^2 f_c^2\right]}. \tag{6}$$

In **Figure 1**, we plot the response of 1st- and 2nd-order APFs based on **Eqs. (2)** and **(4)**.

**Figure 1a** shows the amplitude response, which remains flat over the entire frequency range for both the 1st- and 2nd-order APFs, highlighting a defining feature of these filters. **Figure 1b** shows the phase response of 1st-order APFs with various center frequencies of $f_c$ = 1, 3, and 5 GHz, where the phase decreases monotonically from 0 to -π as the frequency increases, reaching -0.5π at $f_c$. **Figure 1c** shows the group delay response of 1st-order APFs, derived from the phase response in **Figure 1b**, which decreases monotonically with increasing frequency. The maximum group delay is achieved at $f = 0$, and a lower $f_c$ results in a higher maximum group delay.

**Figure 1d** shows the phase response of 2nd-order APFs with various center frequencies (*i.e.*, $f_c$ = 1, 3, and 5 GHz) but the same phase damping factor (*i.e.*, $\xi = 0.1$). Compared with the 1st-order APFs, the 2nd-order APFs exhibit steeper phase transitions around $f_c$ and over a broader range from 0 to -2π. **Figure 1e** shows the group delay response of 2nd-order APFs derived from the phase response in **Figure 1d**. Unlike the 1st-order APFs, the maximum group delay of the 2nd-order APFs is achieved near $f_c$. It is also worth noting that the maximum group delay decreases for increasing $f_c$. This is because, according to **Eq. (6)**, the maximum group delay scales inversely with $\pi\xi f_c$. Therefore, a larger $f_c$ leads to a smaller maximum group delay.

**Figure 1f** shows the phase response of 2nd-order APFs with various phase damping factors (*i.e.*, $\xi$ = 0.1, 0.2, and 0.3) but the same center frequency (*i.e.*, $f_c$ = 5 GHz). As $\xi$ decreases, the phase transition around $f_c$ becomes steeper, leading to a narrower phase bandwidth (PBW), defined as the frequency range over which the phase response varies from -0.5π to -1.5π [21]. **Figure 1g** shows the corresponding group delay response, which exhibits a profile similar to that in **Figure 1e**, with the maximum group delay achieved near $f_c$. As $\xi$ decreases, the maximum group delay increases, corresponding to a faster phase transition and a sharper phase roll-off.

In addition to varying only a single characteristic parameter (as shown in **Figures 1d** and **1f**), simultaneous variation of both characteristic parameters in **Eq. (4)** can realize more advanced phase engineering functionalities. In **Figure 1h,** both $f_c$ and $\xi$ are jointly varied to achieve consistent phase roll-offs with identical maximum group delays across different center frequencies, as required for phase shifters with reconfigurable center frequencies [31, 32]. Here we show the results for three different pairs of ($f_c$, $\xi$), including (1 GHz, 0.1), (3 GHz, 0.033), and (5 GHz, 0.02), and **Figure 1i** shows the corresponding group delay response. As seen, the phase response exhibits consistent roll-offs around different center frequencies, and the group delay response shows identical maximum values across the three cases.

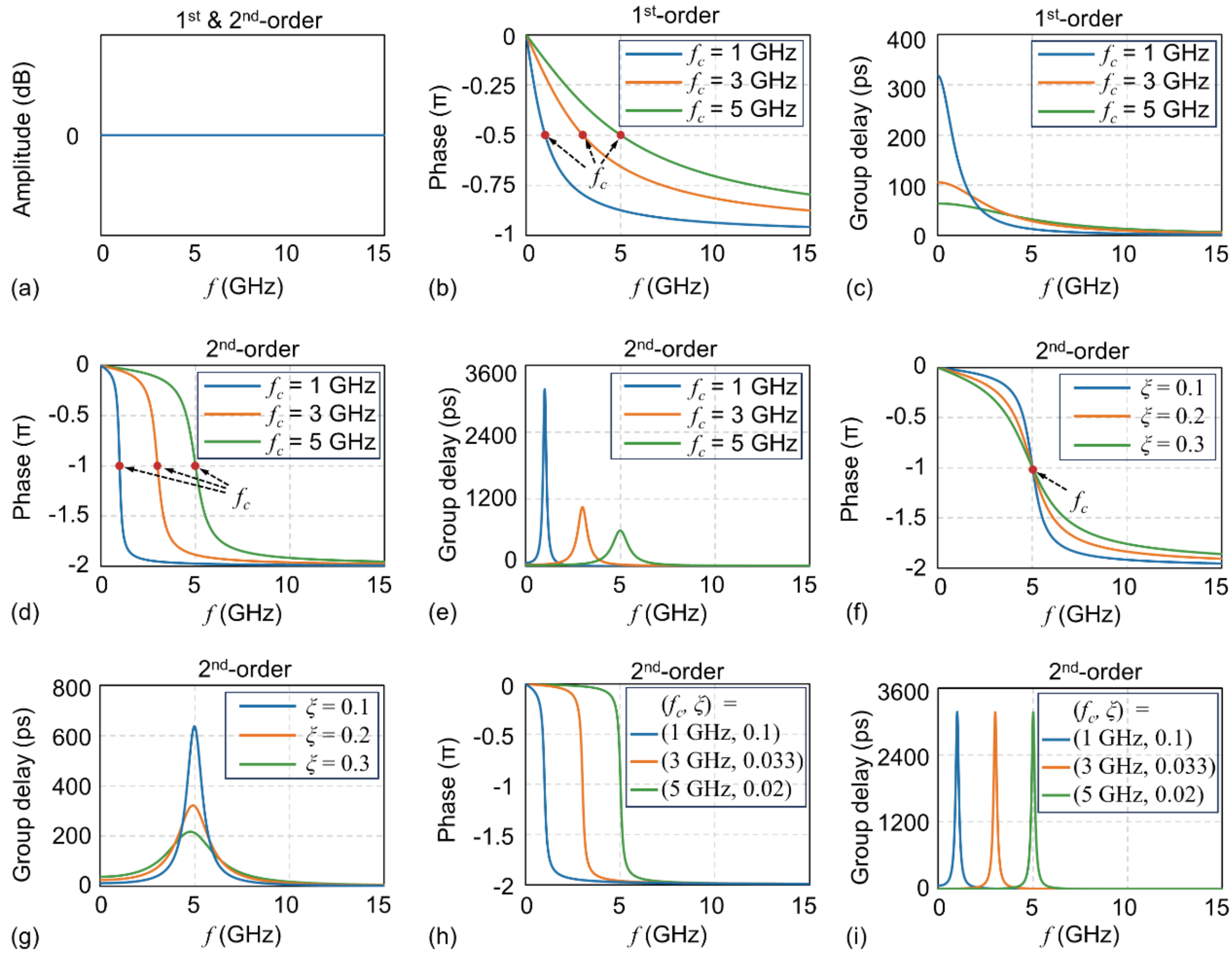

**Figure 1.** Response of ideal all-pass filters (APFs). (a) Amplitude response of 1st- and 2nd-order APFs, where the two response curves overlap with each other. (b) Phase and (c) corresponding group delay response of 1st-order APFs with various center frequencies of $f_c$ = 1, 3, and 5 GHz. (d) Phase and (e) corresponding group delay response of 2nd-order APFs with various $f_c$ = 1, 3, and 5 GHz but the same $\xi$ = 0.1. (f) Phase and (g) corresponding group delay response of 2nd-order APFs with various $\xi$ = 0.1, 0.2, and 0.3 but the same $f_c$ = 5 GHz. (h) Phase and (i) corresponding group delay response of 2nd-order APFs with consistent phase roll-off and identical maximum group delays at $f_c$ = 1, 3, and 5 GHz. The corresponding $\xi$ values are 0.1, 0.033, and 0.02, respectively.

Microwave APFs have been widely used for phase engineering in microwave signal processing, enabling key functionalities such as phase shifters, true time delay lines, and beamforming [9, 20-22, 33, 34]. Reconfigurable microwave APFs, which allow dynamic tuning of either one or both characteristic parameters (*i.e.*, $f_c$ and $\xi$), offer a high degree of flexibility in controlling and engineering the phase response, making them versatile for meeting diverse and evolving requirements.

To realize the APFs shown in **Figure 1**, we employ an MWP transversal filter system, where multiple taps across different channels are linearly combined, making it functionally equivalent to digital signal processing (DSP) filters [35-37]. APFs implemented using MWP technologies and photonic hardware offer several compelling advantages over their electronic counterparts, including wide operating bandwidths that overcome the bandwidth bottleneck of electronic devices, low transmission loss that improves the signal-to-noise ratio, and strong immunity to electromagnetic interference [6, 38, 39].

**Figure 2** illustrates the operation principle and processing flow of a microcomb-based MWP transversal filter system. In this system, optical microcombs generated by a compact micro-resonator provide a multi-wavelength source for constructing wavelength channels corresponding to different taps, enabling a substantial reduction in system size, weight, and power (SWaP) compared with conventional multi-wavelength sources based on discrete laser arrays [40, 41]. As the initially generated comb lines exhibit uneven power levels, the generated microcomb first passes through a flattening stage to equalize the intensities across different wavelength channels. Subsequently, each wavelength channel is assigned a specific weight according to the designed tap coefficient. The tap coefficients are derived from the target APF spectral transfer function via an inverse Fourier transform. In the delay and sum module, replicas of input microwave signal (to be filtered) are generated by modulating it onto comb lines at different wavelength channels via a single electro-optic modulator (EOM). The delay between adjacent channels is introduced by exploiting dispersion of a spool of single-mode fiber (SMF). Finally, the delayed and weighted signal replicas are summed upon photodetection, thereby generating a microwave signal as the filter output.

For the MWP transversal filter system in **Figure 2**, the output microwave signal $s(t)$ can be expressed as

$$s(t) = f(t) * h(t) = \sum_{n=0}^{M-1} a_n f(t - n\Delta t), \tag{8}$$

where $M$ is the tap number (*i.e.*, the number of wavelength channels), $a_n$ ($n$ = 0, 1, 2, …, $M$ - 1) is the tap coefficient of the $n_{th}$ tap, $\Delta t$ is the time delay between adjacent wavelength channels, and $f(t)$ denotes the input microwave signal. The temporal impulse response $h(t)$ in **Eq. (8)** can be given by

$$h(t) = \sum_{n=0}^{M-1} a_n \delta(t - n\Delta t), \tag{9}$$

where $\delta(t)$ is the unit impulse function. By applying the Fourier transform to **Eq. (9)**, the spectral transfer function of the MWP transversal filter system can be written as

$$H(\omega) = \sum_{n=0}^{M-1} a_n e^{-j\omega n\Delta t}, \tag{10}$$

**Eq. (10)** is a typical spectral transfer function for transversal filter systems [42], which exhibits finite impulse response (FIR). By changing the tap coefficients $a_n$ ($n$ = 0, 1, 2, …, $M$ - 1), diverse filter functions can be realized. This forms the basis for realizing the APFs shown in **Figure 1**. Moreover, it enables reconfigurable APF order, center frequency $f_c$, and phase damping factor $\xi$ through programing the tap coefficients $a_n$ ($n$ = 0, 1, 2, …, $M$ - 1).

In the microcomb-based MWP transversal filter system, the comb spacing defines the Nyquist zone, such that the microwave modulation frequency should remain within half of the comb spacing to avoid spectral overlap between adjacent wavelength channels. Once this condition is satisfied, the system operation bandwidth is determined only by the time delay $\Delta t$ between adjacent wavelength channels, which can be calculated as [41]

$$\Delta t = L \cdot D \cdot \Delta\lambda, \tag{11}$$

where $\Delta\lambda$ is the comb spacing, $D$ is the dispersion parameter, and $L$ is the length of the dispersive medium (*i.e.*, the SMF in **Figure 2**). Since the MWP transversal filter has FIR, its microwave spectral response is periodic, with a free spectral range (FSR) given by $FSR_{MW}$ = $1/\Delta t$. The system operation bandwidth, defined as the usable baseband processing range, is typically half of this period, *i.e.*, $FSR_{MW}$ / 2 = 1 / (2$\Delta t$).

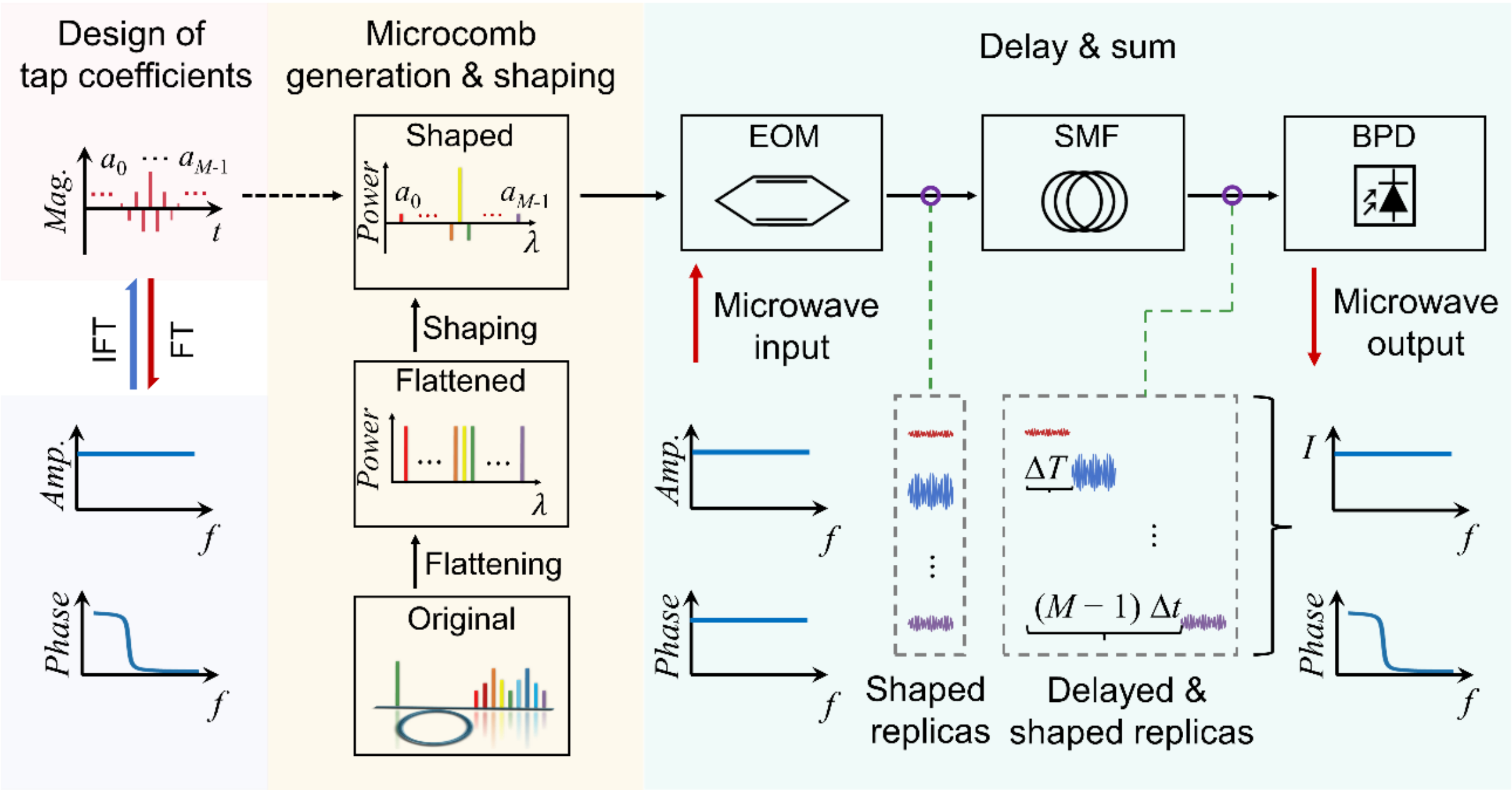

**Figure 2.** Schematic diagram and processing flow of a microwave photonic (MWP) transversal filter system with an optical microcomb source, where $\Delta t$ is the time delay between adjacent wavelength channels, and $a_0$, …, $a_{M-1}$ are the tap coefficients for different wavelength channels. IFT: inverse Fourier transform. FT: Fourier transform. EOM: electro-optic modulator. SMF: single-mode fiber. BPD: balanced photodetector.

**Figure 3** shows the simulated response of APFs implemented using the microcomb-based MWP transversal filter system in **Figure 2**. In our simulation, the system bandwidth was first determined by **Eq. (11)**, and the target transfer functions in **Eqs. (2)** and **(4)** were evaluated over the corresponding frequency grid. The impulse response was then obtained via inverse Fourier transform and discretized into a finite number of taps, with the zero-delay tap defined as the center tap and the remaining taps selected symmetrically on both sides, thereby generating the tap coefficients in **Eq. (10)**. The characteristic parameter for 1$^{st}$-order APFs was set to $f_c$ = 1 GHz, and the characteristic parameters for 2$^{nd}$-order APFs were set to $f_c$ = 5 GHz and $\xi$ = 0.15. The system operation bandwidth was set to ~15 GHz, consistent with that used in the subsequent experimental demonstration.

**Figure 3a** and **3b** show the designed tap coefficients for 1$^{st}$- and 2$^{nd}$-order APFs, respectively, each providing the results for four different tap numbers of $M$ = 11, 21, 41 and 81. For 1$^{st}$-order APFs, all tap coefficients on the right-hand side of the center are negative, with their magnitudes decreasing as the distance from the center increases, whereas those on the left-hand side remain close to zero. For 2$^{nd}$-order APFs, the tap coefficients on the left-hand side also remain close to zero – similar to those of the 1$^{st}$-order APFs. On the right-hand side, both positive and negative tap coefficients are present, with their magnitudes decreasing as the distance from the center increases, indicating that taps near the center play a dominant role in response reconstruction. The increased complexity of tap coefficient distributions for 2$^{nd}$-order APFs stems from their more complex phase response.

**Figure 3c** shows the simulated amplitude and phase response of the APFs based on the designed tap coefficients in **Figures 3a** and **3b**. For both 1$^{st}$- and 2$^{nd}$-order APFs, increasing the tap number leads to amplitude and phase response more closely approaching that of the ideal APFs in **Figure 1**. This trend arises from the FIR nature of the transversal filter systems, which introduce deviations from the ideal response when approximating target filters with infinite impulse response (IIR), such as the APFs investigated in this work. The deviations between the ideal response and that realized based on **Eq. (10)**, also referred to as theoretical approximation errors, decreases as the tap number increases, as the IIR system has more feedback paths for adjustment.

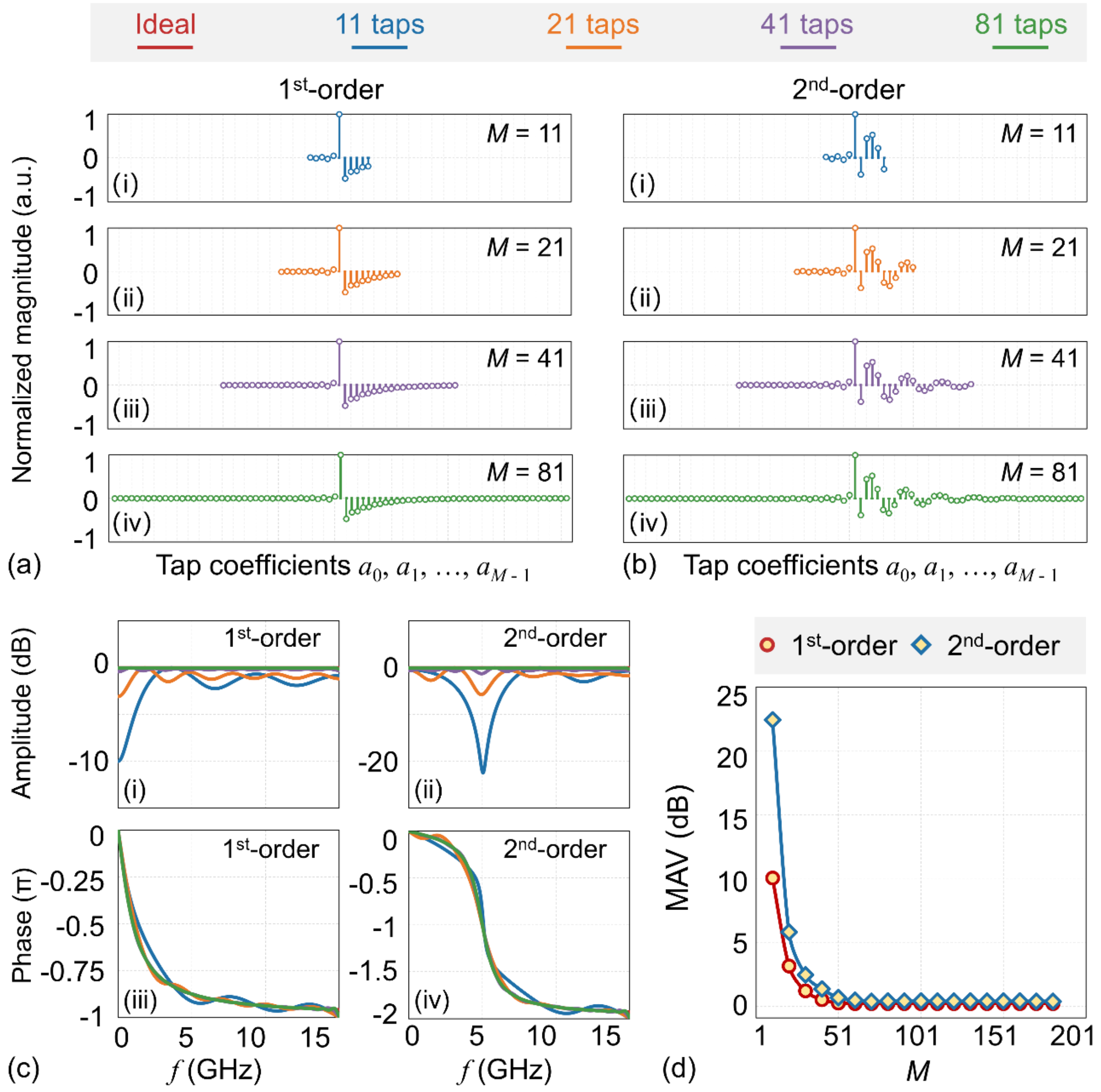

**Figure 3.** Simulated response of APFs realized using a microcomb-based MWP transversal filter system. (a) – (b) Designed tap coefficients for 1st- and 2nd-order APFs, respectively. In each figure, (i) – (iv) show the results for various tap numbers of $M$ = 11, 21, 41, and 81, respectively. (c) Simulated amplitude and phase response for 1st- and 2nd-order APFs in (a) and (b), respectively. (d) Maximum amplitude variation (MAV) versus tap number ($M$).

To quantify the deviations between the transversal filter response and the ideal APF response, we introduce a parameter termed maximum amplitude variation (MAV), defined as the maximum deviation of the amplitude response from the ideal flat amplitude response over the frequency range of interest. **Figure 3d** shows the MAV versus tap number $M$. For both 1st- and 2nd-order APFs, the MAV decreases rapidly with increasing $M$, indicating a substantial improvement in approximation accuracy as the tap number increases. At small tap numbers, the 2nd-order APF exhibits a larger MAV than the 1st-order APF, reflecting its higher sensitivity to tap number due to its more complex phase response. In addition, the MAV reaches a sufficiently low level at $M = 41$ for both types of APFs, beyond which further increase in the tap number yields only marginal improvement. This indicates that 41 tap coefficients are sufficient for accurate response synthesis in the transversal filter system. Considering this, a maximum tap number of $M = 41$ is chosen for our following experimental demonstrations.

**Figure 4** shows the simulated response of APFs with reconfigurable characteristic parameters. In each of **Figures 4a – d**, we show the results for **(i)** amplitude response, **(ii)** MAV versus $f_c$, **(iii)** phase response, and **(iv)** group delay response. In our simulation, the tap number is fixed at $M = 41$ for comparison, and the system operation bandwidth was set to 15 GHz, the same as in **Figure 3** and in our following experimental demonstration.

**Figure 4a** shows the results for a 1st-order APF with reconfigurable center frequency of $f_c = 1 – 11$ GHz. In all cases, the amplitude response remains close to 0 dB, as expected for APFs. In addition, the MAV characterizing the amplitude ripples exhibits a strong dependence on $f_c$. With increasing $f_c$, the MAV decreases initially and subsequently increases, achieving a minimum value at an intermediate $f_c$. This is because a small $f_c$ leads to a steep phase roll-off, increasing theoretical approximation errors using **Eq. (10)** and thus the MAV. A large $f_c$ approaching the system operation bandwidth also increases the MAV, due to the finite operation bandwidth of the transversal filter system (*i.e.*, 15 GHz in our simulation). The phase and group delay response also closely matches that of the ideal 1st-order APFs in **Figures 1(b)** and **(c)**, respectively. A higher $f_c$ leads to a more gradual phase roll-off and thus a decreased maximum group delay at $f = 0$.

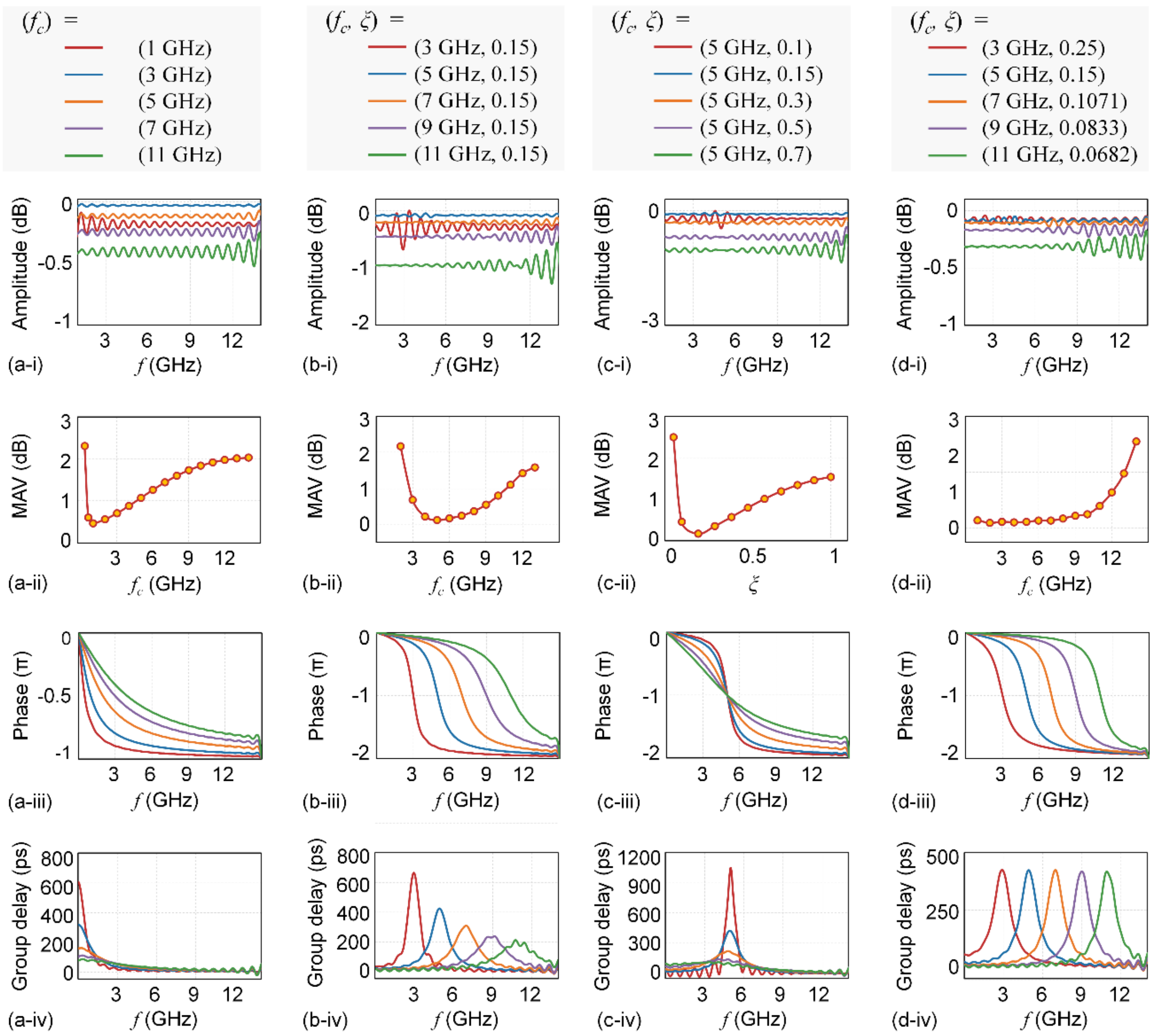

**Figure 4.** Simulated response of APFs for reconfigurable center frequencies ($f_c$) and phase damping factors ($\xi$), realized using the microcomb-based MWP transversal filter system. (a) Simulated response of a 1st-order APF with reconfigurable $f_c$ = 0.5, 1, 2, 3, and 4 GHz. (b) Simulated response of a 2nd-order APF with reconfigurable $f_c$ = 3, 5, 7, 9, and 11 GHz at a fixed $\xi$ = 0.15. (c) Simulated response of a 2nd-order APFs with reconfigurable $\xi$ = 0.1, 0.15, 0.3, 0.5, and 0.7 at a fixed $f_c$ = 5. (d) Simulated response of a 2nd-order APFs with consistent phase roll-off and identical maximum group delays at $f_c$ = 3, 5, 7, 9, and 11 GHz, with the corresponding $\xi$ of 0.25, 0.15, 0.1071, 0.0833, and 0.0682, respectively. In (a) – (d), (i) and (iii) show the amplitude and phase response, respectively, and (ii) and (iv) show the MAV and group delay calculated from (i) and (iii), respectively.

**Figure 4b** shows the results for a 2nd-order APF with reconfigurable $f_c$ = 3 – 11 GHz at a fixed phase damping factor of $\xi$ = 0.15. Similar to the observation for the 1st-order APF, the amplitude response remains close to 0 dB. In **Figure 4b-ii**, the MAV is relatively large at a small $f_c$, which can be attributed to the steeper phase transition induced by a fixed $\xi$ at a lower $f_c$. As $f_c$ further increases, the MAV rises again when $f_c$ approaches the system operation bandwidth. In **Figure 4b-iii**, increasing $f_c$ shifts the phase transition toward higher frequencies and makes the phase roll-off more gradual, resulting in decreased maximum group delay with increasing $f_c$ in **Figure 4b-iv**. These trends are also consistent with those in **Figures 1d** and **1e**, respectively.

**Figure 4c** shows the results for a 2nd-order APF with reconfigurable $\xi$ = 0.1 – 0.7 at a fixed $f_c$ = 5 GHz. Relatively significant ripples are observed for $\xi$ = 0.1, resulting in a relatively high MAV. This is mainly attributed to the increased amplitude variation required to realize a faster phase transition at a small $\xi$. When $\xi$ becomes large, the phase roll-off becomes overly gradual, leading to a phase jump near the system operation bandwidth, thereby increasing the amplitude variation and hence the MAV. The phase and group delay responses in **Figures 4c-iii** and **4c-iv** also show good agreement with those in **Figures 1f** and **1g**, respectively.

**Figure 4d** shows the results for a 2nd-order APF with both reconfigurable $f_c$ and $\xi$ to achieve consistent phase-roll-offs with identical maximum group delays. Here we show the results for five different pairs of ($f_c$, $\xi$), including (3 GHz, 0.25), (5 GHz, 0.15), (7 GHz, 0.1071), (9 GHz, 0.0833), and (11 GHz, 0.0682). The MAV remains nearly unchanged at small $f_c$ values and increases more significantly at larger $f_c$ values. This trend arises from the nearly constant phase roll-off, such that significant errors occur only when $f_c$ approaches the system operation bandwidth. The phase response in **Figure 4d-iii** exhibits consistent roll-offs across different $f_c$, and the group delay response in **Figure 4d-iv** shows identical maximum values. Both are in good agreement with those in **Figures 1h** and **1i**, respectively.

The results in **Figure 4** highlight the high degree of reconfigurability of MWP APFs realized using a microcomb-based transversal filter system, where flexible tuning of the characteristic parameters, and even switching between 1st- and 2nd-order response, can be achieved by simply programming the tap coefficients, without any changes to the physical hardware. As shown in **Figures 4b – d**, for 2nd-order APFs with enhanced phase engineering capability, the two characteristic parameters (*i.e.*, $f_c$ and $\xi$) can be tuned either independently or jointly to realize different functionalities, greatly expanding their versatility in practical applications. In contrast, APFs implemented using integrated passive filters such as MRRs [20, 21] typically have limited reconfigurability, as their characteristic parameters are jointly determined by multiple structural parameters [43, 44], restricting both independent and joint tuning over a wide parameter space, although minor adjustments can be achieved using thermal heaters.

## ■ EXPERIMENTAL SETUP

Based on the theory detailed in previous section, we performed experimental demonstrations for reconfigurable MWP APFs. **Figure 5** shows a schematic of the experimental setup. A tunable continuous-wave (CW) laser was first amplified by an erbium-doped fiber amplifier (EDFA), followed by polarization adjustment using a polarization controller (PC). The amplified pump light was coupled into an integrated MRR, which was regulated by a temperature controller (TC), thereby enabling microcomb generation with long-term stability. Since the initially generated microcomb exhibited non-uniform comb line powers, an optical spectral shaper (OSS) was employed to pre-shape the comb spectrum, resulting in a flattened output from the microcomb generation module.

The flattened microcomb was amplified by another EDFA and then sent into the transversal filter module, which consisted of a PC, an EOM, a spool of SMF, a second OSS, and a balanced photodetector (BPD). The PC was used to optimize the modulation efficiency of the polarization-sensitive EOM. The EOM, SMF, second OSS, and BPD performed the same functions as discussed in **Figure 2**. The two complementary outputs of the OSS were fed into the BPD, separating the wavelength channels into two sets. Balanced photodetection introduces a π phase difference between the two sets, thereby enabling both positive and negative tap coefficients. To minimize the time-delay mismatch between the two tap groups, a length-matched fiber pair was incorporated into the optical paths before the BPD.

Except for the microcomb-based transversal filter system, an optical spectrum analyzer (OSA, Anritsu) was employed to characterize the comb spectrum, and a vector network analyzer (VNA, Rohde & Schwarz) was used to characterize the system's spectral response by sweeping the microwave frequency. To mitigate errors arising from the non-ideal response of experimental components and improve the accuracy of comb line shaping, a two-stage feedback control scheme was also implemented, incorporating coordinated synergic spectral power reshaping and impulse response reshaping, as detailed in Ref. [45].

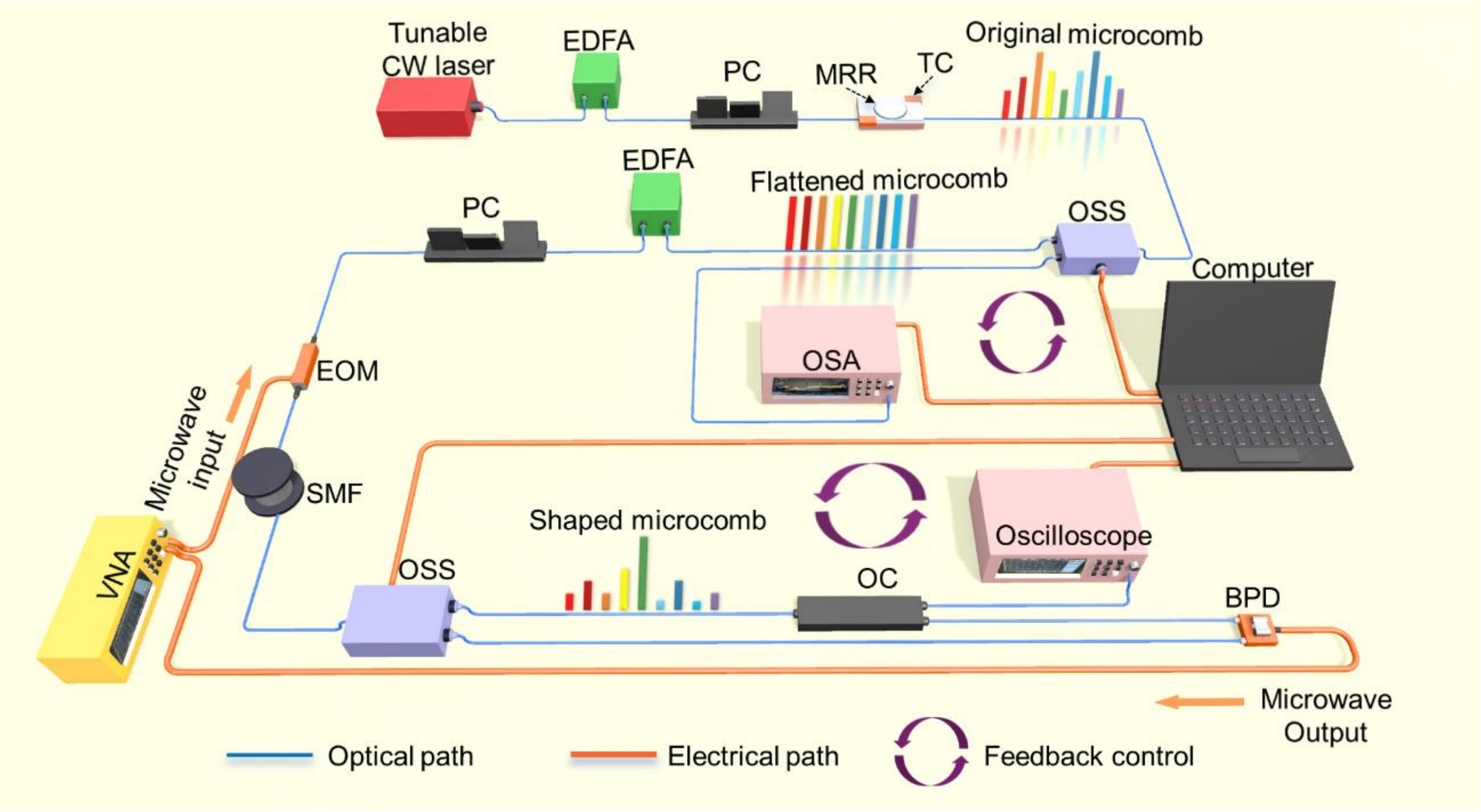

**Figure 5.** Schematic illustration of experimental setup for demonstrating reconfigurable microcomb-based MWP APFs. CW laser: continuous-wave laser. EDFA: erbium-doped fiber amplifier. PC: polarization controller. MRR: micro-ring resonator. TC: temperature controller. OSS: optical spectral shaper. OSA: optical spectrum analyzer. EOM: electro-optic modulator. SMF: single-mode fiber. OC: optical coupler. BPD: balanced photodetector. VNA: vector network analyzer.

In our experimental demonstration, we employed an integrated doped silica MRR to generate optical microcombs as the source for the transversal filter system. **Figure 6a** illustrates the microcomb generation process, in which a high-power continuous-wave (CW) pump is coupled into the ring resonator, producing a number of comb lines at the output through nonlinear interaction within the cavity.

**Figure 6b** shows a microscope image of a doped silica MRR, which was fabricated using complementary metal-oxide-semiconductor (CMOS) compatible fabrication processes [46, 47]. During the fabrication, doped silica films with a refractive index of ~1.7 at 1550 nm were deposited by plasma-enhanced chemical vapor deposition (PECVD), followed by patterning through deep ultraviolet (DUV) photolithography and reactive-ion etching to form low-loss waveguides. After this, a silica upper cladding with a refractive index of ~1.44 at 1550 nm was deposited. Finally, the integrated chip is packaged with mode convertors that connect to fiber-pigtails, with a coupling loss below 1 dB per facet. Th doped silica platform features a low linear propagation loss of ~0.06 dB · $cm^{-1}$, a moderately high nonlinear parameter of ~233 $W^{-1}$ · $km^{-1}$, and negligible nonlinear loss up to intensities of ~25 GW · $cm^{-2}$ [48]. The fabricated MRR exhibited a high-quality (Q) factor of ~1.9 million and an FSR ~0.4 nm (*i.e.*, ~49 GHz).

**Figure 6c** shows the optical spectrum of the soliton crystal (SC) microcomb generated by the MRR in **Figure 6b**, which has a comb spacing that equals the FSR of the MRR. To generate the microcomb, a CW pump was amplified to ~32.1 dBm and swept from shorter to longer wavelengths across a TE-polarized resonance near 1551.3 nm. As the pump approached the cold-cavity resonance, the intracavity power increased and eventually exceeded the threshold for modulation instability (MI), thereby initiating MI oscillation [49]. This first led to the emergence of primary comb lines, whose initial spacing was determined by the MI gain peak governed mainly by the cavity dispersion and intracavity power. With further increase in pump detuning, the spectrum evolved into the characteristic fingerprint-like profile of a soliton-crystal state, consistent with previous reports in Refs. [50-52].

Unlike conventional single soliton states, SC microcombs arise from the self-organization of multiple co-circulating solitons into an ordered intracavity configuration within the MRR [53, 54]. Compared with dissipative Kerr solitons [7, 41], SC microcombs exhibit only minimal variation in intracavity energy during formation, which allows simple and robust initiation through adiabatic manual tuning of the pump wavelength [7, 54]. An enlarged view of the SC microcomb spectrum within the telecommunications C band (*i.e.*, 1530-1565 nm) is shown in **Figure 6d**. Owing to the relatively small FSR of the MRR, 90 comb lines are generated within the C band, providing sufficient taps for use in the transversal filter system.

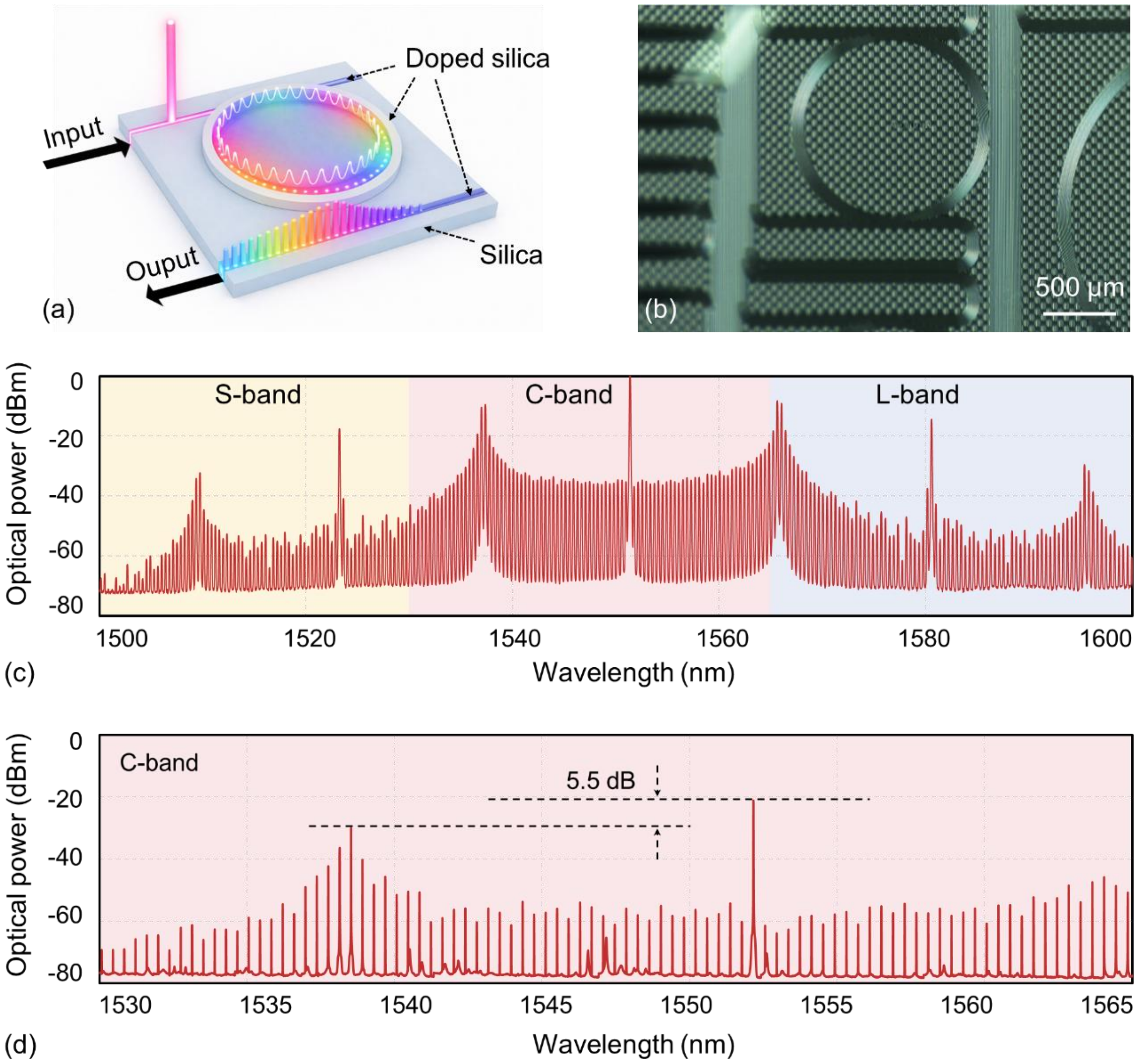

**Figure 6.** Optical microcomb generation. (a) Schematic illustration of optical microcomb generation in a doped silica micro-ring resonator (MRR). (b) Microscope image of a fabricated doped silica MRR. (c) Measured optical spectrum of soliton-crystal microcomb generated by the MRR in (b), covering S-, C-, and L- bands. (d) Zoom-in view of the spectrum in (c) within the C-band.

For the experimental setup shown in **Figure 5**, the SMF had a length of 4.8 km and a dispersion of 17.4 ps · nm$^{-1}$ · km$^{-1}$. Combined with the microcomb spacing of ~0.4 nm in **Figure 6c**, this yields a time delay of ~0.033 ns between adjacent channels, corresponding to a microwave FSR of ~30 GHz calculated based on **Eq. (11)**. Therefore, the operation bandwidth of the practical system in our experimental demonstration is ~30/2 = ~15 GHz.

## EXPERIMENTAL RESULTS

By using the experimental setup discussed in previous section, we performed demonstrations for reconfigurable MWP APFs. In this section, we present and discuss the experimental results, including the filter response for various tap numbers and with reconfigurable characteristic parameters.

**Figure 7** shows the experimental results for 1st-order APFs with various tap numbers of $M$ = 11, 21, and 41. For comparison, the center frequency was set to $f_c$ = 1GHz. **Figures 7(a-i) – (a-iii)** show the designed tap coefficients and the measured optical spectra of the shaped comb lines for $M$ = 11, 21, and 41, respectively. The circles denote the target tap coefficients, and the solid lines represent the experimentally shaped comb spectra, with yellow and orange corresponding to the positive and negative taps, respectively. With increasing tap number, more comb lines are employed in the transversal filter, resulting in a closer approximation to the ideal APF transfer function in **Eq. (2)**.

**Figure 7b** shows the measured amplitude response of the APFs. It can be seen that increasing the tap number leads to a substantial improvement in the response flatness, with the measured amplitude response exhibiting reduced ripples, particularly at low frequencies. These phenomena are consistent with the simulation results in **Figure 3c-i**. **Figure 7c** shows the MAV versus $M$ calculated based on the results in **Figure 7b**. As the tap number increases from 11 to 41, the MAV decreases from ~13.03 dB to ~1.95 dB. reflecting the effective suppression of amplitude ripples with increasing $M$.

**Figure 7d** shows the measured phase response of the APFs. The phase response decreases from 0 to -π, and becomes closer to the ideal phase response of 1st-order APFs as $M$ increases. These are consistent with the simulation results in **Figure 3c-iii**. **Figure 7e** shows the group delay response calculated from the phase response in **Figure 7d**. The maximum group delays are observed in the low frequency region, followed by a rapid decay with increasing frequency, in good agreement with the simulation results in **Figure 4a-iv**.

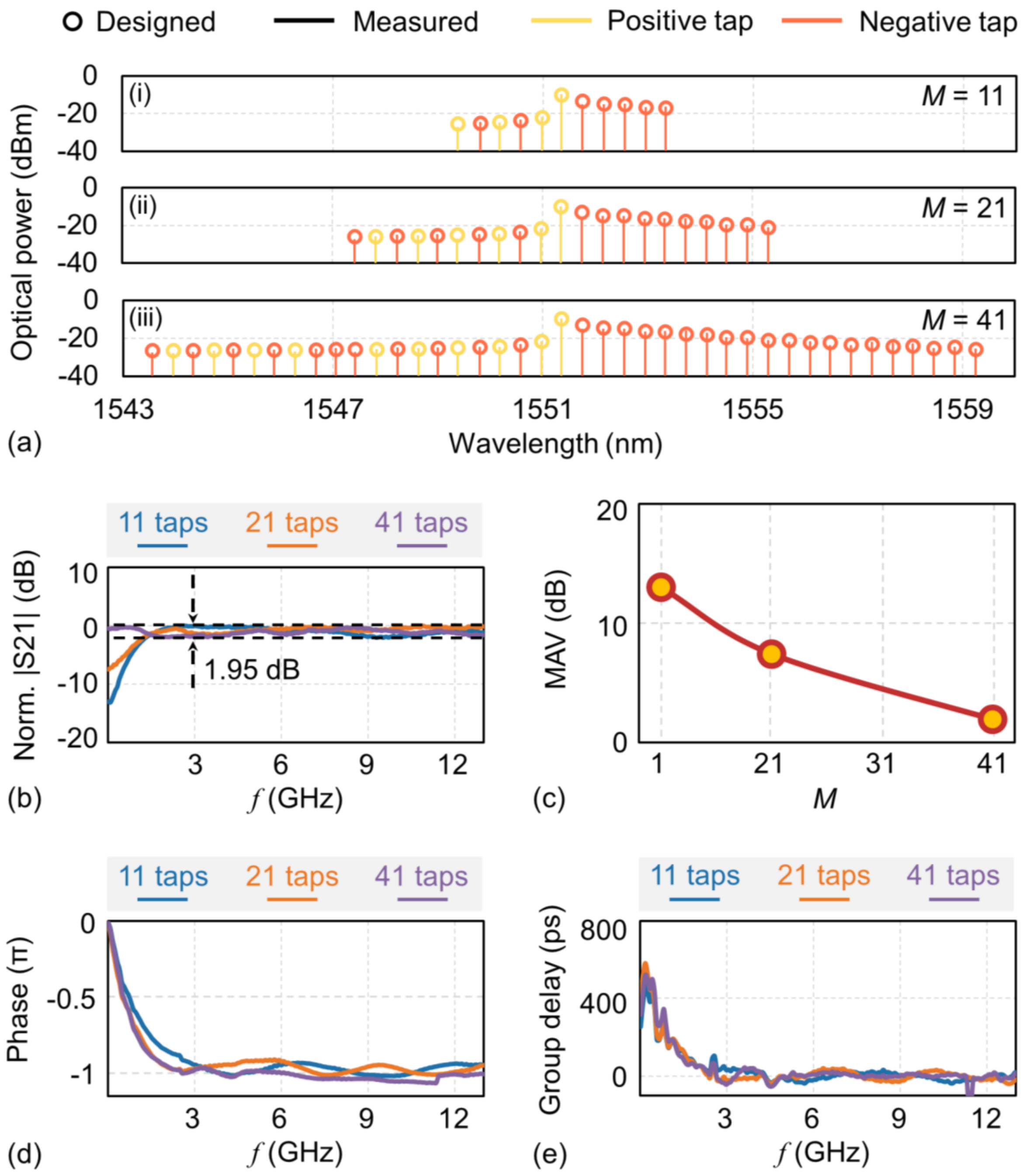

**Figure 7.** Experimental demonstration for 1st-order MWP APFs with various tap numbers of $M$ = 11, 21, and 41. (a) Designed tap coefficients (circles) and measured optical spectra of shaped comb lines (solid lines) for (i) $M$ = 11, (ii) $M$ = 21, and (iii) $M$ = 41, where yellow circles and lines represent positive tap coefficients, and orange circles and lines correspond to negative tap coefficients. (b) Measured amplitude response corresponding to the results in (a). (c) MAV versus $M$ calculated based on the results in (b). (d) Measured phase response corresponding to the results in (a). (e) Group delay calculated based on the results in (d). In (a) – (e), the center frequency is $f_c$ = 1 GHz.

**Figure 8** shows the experimental results for $2^{nd}$-order APFs with various tap numbers of $M$ = 11, 21, and 41. For comparison, the center frequency and the phase damping factor were set to $f_c$ = 5 GHz and $\xi$ = 0.15, respectively. **Figure 8a** shows the designed tap coefficients and the measured optical spectra of the shaped comb lines, where the symbols and colors have the same meanings as those in **Figure 7a**, and a larger $M$ allows more accurate approximation of the target transfer function in **Eq. (4)**. **Figure 8b** shows the measured amplitude response of the APFs. Similar to that observed in **Figure 7b**, increasing the tap number improves the response flatness and reduces the amplitude ripples, particularly around $f_c$.

**Figure 8c** shows the MAV versus $M$ calculated based on the results in **Figure 8b.** As the tap number increases from 11 to 41, the MAV decreases monotonically from ~33.6 dB at $M$ = 11 to ~2.01 dB at $M$ = 41. The phase response in **Figure 8d** decreases from 0 to -2π, and becomes closer to the ideal phase response of $2^{nd}$-order APFs as $M$ increases. For $M$ = 11, 21, and 41, the $PBW$ are ~1.28 GHz, ~1.92 GHz, and ~1.5 GHz, respectively. **Figure 8e** shows the group delay response calculated from the phase response in **Figure 8d**, where the maximum group delays are achieved near $f_c$. A higher maximum group delay is achieved for a smaller $M$, mainly arising from the more significant amplitude ripples under this condition.

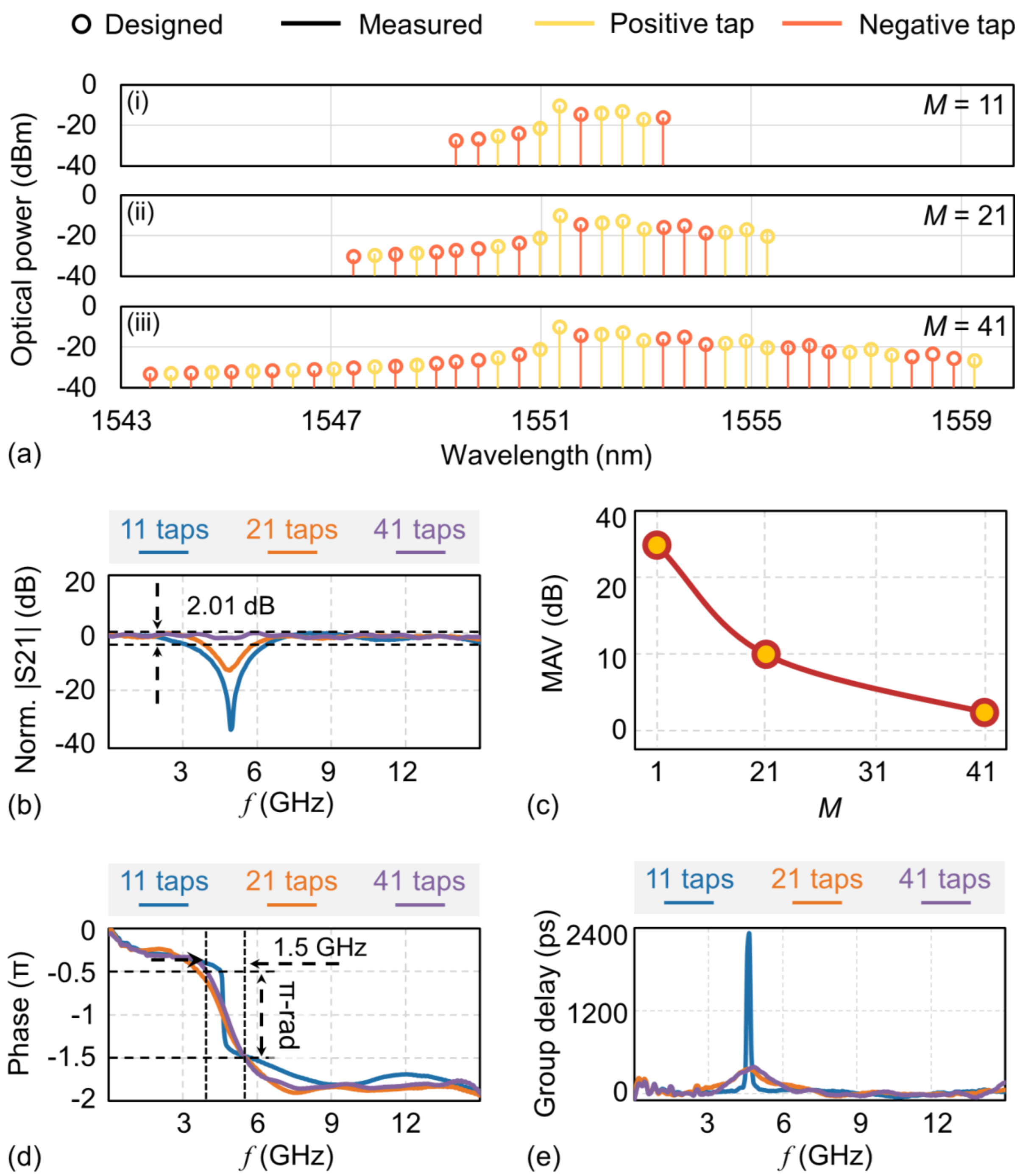

**Figure 8.** Experimental demonstration for 2nd-order MWP APFs with various tap numbers of $M = 11$, 21 and 41. (a) Designed tap coefficients (circles) and measured optical spectra of shaped comb lines (solid lines) for (i) $M = 11$, (ii) $M = 21$, and (iii) $M = 41$, where yellow circles and lines represent positive tap coefficients, and orange circles and lines correspond to negative tap coefficients. (b) Measured amplitude response corresponding to the results in (a). (c) MAV versus $M$ calculated based on the results in (b). (d) Measured phase response corresponding to the results in (a). (e) Group delay calculated based on the results in (d). In (a) – (e), the center frequency and the phase damping factor are $f_c = 5$ GHz and $\xi = 0.15$, respectively.

**Figure 9** shows experimental results of APFs with reconfigurable characteristic parameters. In each of **Figures 9a – d**, we show the results for **(i)** amplitude response, **(ii)** phase response, and **(iii)** group delay response.

**Figure 9a** shows the results for a 1st-order APF with reconfigurable center frequency of $f_c$ = 0.5, 1, and 3 GHz. The amplitude response remains relatively flat in all cases, and the MAV is ~ 2.98 dB. The phase response exhibits the expected monotonic decrease for various $f_c$, in good agreement with the simulation results in **Figure 4a-iii.** The corresponding group delay response reaches a maximum in the low frequency region and then gradually decreases with increasing frequency, consistent with the results in **Figure 4a-iv**. At $f_c$ = 0.5 GHz, the filter achieves an MGD of ~580 ps near $f$ = 0 GHz.

**Figure 9b** shows the results for a 2nd-order APF with reconfigurable $f_c$ = 5, 7 and 9 GHz at a fixed phase damping factor of $\xi$ = 0.15. The amplitude response also preserves a high degree of flatness in all cases. The MAV is ~4.36 dB, which is slightly higher than that in **Figure 9a** due to the more complex tap-coefficient distribution of the 2nd-order APF, consistent with the simulation results shown in **Figures 3c** and **3d**. As $f_c$ increases, the phase transition shifts toward higher frequencies, accompanied by a decrease in the maximum group delay, showing agreement with the simulation results in **Figure 4b**.

**Figure 9c** shows the results for a 2nd-order APF with reconfigurable $\xi$ = 0.05, 0.15, and 0.3 at a fixed $f_c$ = 5 GHz. A narrow *PBW* of ~0.5 GHz, corresponding to an MGD of ~1740 ps, is achieved for $\xi$ = 0.05, together with relatively large ripples that result in a MAV of ~5.97 dB. These results reflect the trade-off between achieving sharp phase transition and maintaining high amplitude-response flatness. As $\xi$ increases, the phase transition becomes more gradual, and the corresponding maximum group delay near $f_c$ decreases with increasing $\xi$, consistent with the simulation results in **Figure 4c**.

**Figure 9d** shows the results for a 2$^{nd}$-order APF with both reconfigurable $f_c$ and $\xi$ to achieve consistent phase-roll-offs with identical maximum group delays. In all cases, only minor amplitude variations are observed, with a MAV of ~4.02 dB. The phase transition shifts toward higher frequencies while preserving a nearly identical phase roll-off. This results in a shift of the group delay peak with little change in its maximum value, which remains ~400 ps. These results confirm reconfigurable tuning of the center frequency $f_c$ with minimal change to the phase-transition characteristic, agreeing well with the simulation results in **Figure 4d**.

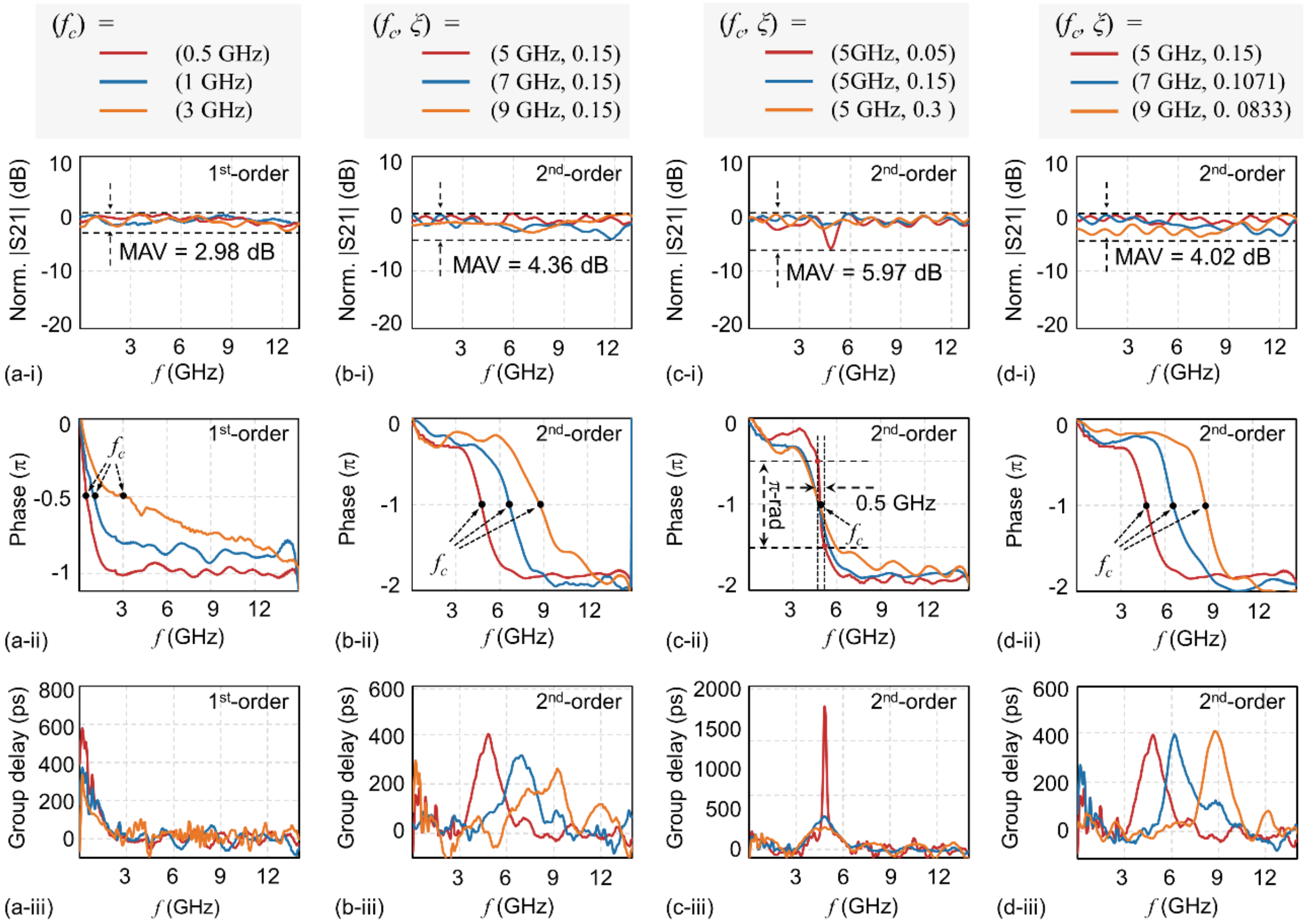

**Figure 9.** Experimental demonstration for reconfigurable MWP APFs. (a) a 1$^{st}$-order APF with reconfigurable $f_c$ = 0.5, 1, and 3 GHz. (b) a 2$^{nd}$-order APF with reconfigurable $f_c$ = 5, 7, and 9 GHz at a fixed $\xi$ = 0.15. (c) a 2$^{nd}$-order APF with reconfigurable $\xi$ = 0.05, 0.15, and 0.3 at a fixed $f_c$ = 5. (d) a 2$^{nd}$-order APF with consistent phase roll-off and identical maximum group delays at $f_c$ = 5, 7, and 9 GHz. The corresponding $\xi$ values are 0.15, 0.1071, and 0.0833, respectively. In (a) – (d), (i), (ii), and (iii) show the amplitude, phase, and group delay response, respectively.

## ■ DISCUSSION

In **Table 1**, we compare our MWP APFs with previous reported implementations in terms of key performance metrics, including MAV, PBW, MGD, insertion loss, and reconfigurability. As seen, previous works have relied on passive optical filters such as MRRs, whereas our work represents the first demonstration of MWP APFs based on optical microcombs.

**Table 1. Comparison of MWP APFs.** MAV: maximum amplitude variation. PBW: phase bandwidth, IL: insertion loss, MRR: micro-ring resonator, MZI: Mach–Zehnder interferometer, MMI: multimode interferometer. MGD: maximum group delay. GD: group delay. CF: center frequency.

| Key device | MAV (dB) | PBW (GHz) | MGD (ps) [b] | IL (dB) [c] | Reconfigurable | | | Ref. |
|---|---|---|---|---|---|---|---|---|
| | | | | | order [e] | GD [e] | CF [e] | |
| MRR | 1.2 | 1.04 | N/A[a] | 8.4 | No | No | Yes | [21] |
| Dual-injection MRR | 0.5 | – [a] | 200 | 5.0 | No | No | Yes | [20] |
| MZI-assisted MRRs | 1.9 | – [a] | 429 | 7.2 | No | Yes | Yes | [33] |
| Dual-injection MRR with MMI | 1.3 | – [a] | 399 | 8.0 | No | No | Yes | [22] |
| Microcombs | 1.95 | 0.50 | 1740 | ~0[d] | Yes | Yes | Yes | This work |

[a] There is no reported value for this parameter in the literature.
[b] Here we show the maximum group delay values achieved in experimental demonstrations.
[c] Here the insertion loss indicates the power loss of the optical filtering module.
[d] Here, a multi-channel transversal filter system is employed. The scheme differs from all the other works listed in Table 1, which are based on optical filtering at a single resonance. In the transversal filter system, negligible IL is introduced by single-mode fibers used as the dispersive module along the optical path.
[e] Here, 'Yes' and 'No' indicate whether reconfigurable APF order, group delay, and center frequency were experimentally demonstrated.

The MAV of our APFs is slightly higher than those reported in previous works. This arises from both theoretical approximation errors inherent to the FIR transversal filter and non-ideal response of practical experimental components. The former were effectively minimized by using a sufficiently high tap number $M$ (up to 41, as discussed in **Fig. 3d**). As a result, the latter, including microcomb noise, EOM chirp, high-order dispersion of the SMF, OSS shaping errors, and BPD noise, play a dominant role. Previously [55, 56], we demonstrated that the static and slowly varying errors arising from the non-ideal response of the EDFA, OSSs, EOM, SMF, and BPD can be mitigated by introducing feedback control into the transversal filter system (as illustrated in **Fig. 5** and adopted in our experimental demonstrations in **Figs. 7 – 9**). However, this approach is less effective for rapidly varying error sources, particularly microcomb and BPD noise. Consequently, residual discrepancies remain between the ideal and experimental responses, which could be further suppressed through advanced mode-locking strategies for microcomb stabilization [57] and gradient-descent-based optimization for enhanced calibration accuracy [58].

Our APFs achieve narrower *PBW*s (*i.e.*, sharper phase transitions) and higher group delays than those reported in previous works. This is enabled by the ease of realizing high-order filter response in the transversal filter architecture. As demonstrated in Ref. [7], our system inherently avoids the resonance wavelength alignment required in APFs based on multiple MRRs [59], as well as the associated performance degradation caused by thermal-drift-induced misalignment. Moreover, the soliton-crystal microcomb employed in this work has demonstrated stable operation for ~66 h in optical communication experiments [60]. Except for the microcomb source, all other components are commercially available and have demonstrated high stability, which further supports the long-term stable operation of the proposed APFs.

As shown in **Table 1**, the previous works based on integrated passive filters, such as MRRs, typically suffer high insertion loss, mainly due to the use of multiple Mach-Zehnder interferometers (MZIs). For our APFs based on a multi-wavelength MWP transversal filter system, the operating mechanism is fundamentally different from that of previous works based on optical filtering at a single resonance, yielding only negligible insertion loss introduced by SMFs used as the dispersive module along the optical path.

Finally, our APFs provide the highest degree of reconfigurability among all the works summarized in **Table 1**. Unlike the previous works, the filter order, group delay, and center frequency of our APFs can all be flexibly reconfigured simply through programming the tap coefficients via the OSS in **Figure 5**, without any changes to the hardware. As demonstrated in **Figure 9**, all the APF characteristic parameters can be tuned either independently or jointly to realize diverse phase engineering functionalities, greatly expanding the versatility of our APFs for practical applications requiring highly reconfigurable phase and delay control, such as optical phased arrays and phased array radars. It is also worth noting that, in principle, all components of the microcomb-based transversal filter system can be integrated on a single chip, although replacing discrete laser array with an integrated microcomb source alone already offers significant benefits in terms of SWaP, cost, and complexity. Recently, exciting progress has been achieved in the integrated implementation of key system components, including microcomb generators [61], optical amplifiers [62, 63], spectral shapers [64, 65], EOMs [66], delay lines [67], and photodetectors [68], paving the way for realizing monolithically integrated microcomb-driven MWP APFs. This work has broad implications for microcombs [69-96] and their applications to microwave photonics, neuromorphic processors and communications. [97-152] The addition and use of 2D materials [153-194] will add extra functionality to microcomb chips for potential applications to quantum photonics [195 - 210] and other areas. [211-230]

## ■ CONCLUSION

In summary, we propose and experimentally demonstrate highly reconfigurable MWP APFs using a microcomb-driven transversal filter system. By leveraging the large number of comb lines provided by optical microcombs, the transversal filter system can synthesize APF response with a high degree of reconfigurability, sharp phase transitions with high group delays, and low loss. We experimentally demonstrated both $1^{st}$- and $2^{nd}$-order APFs, achieving a MAV of ~1.95 dB and an MGD of ~580 ps for the former, and a MAV of ~2.01 dB and an MGD of ~1740 ps for the latter. In addition, the filter response is highly reconfigurable through simply programming the tap coefficients without changing any hardware, allowing both independent and joint tuning of APF characteristic parameters for diverse phase-engineering functionalities. This work provides a new route towards realizing highly reconfigurable, high-performance MWP APFs for practical applications requiring dynamic and versatile phase control.

## ■ REFERENCES